\documentstyle[twoside,fleqn,espcrc2,epsf]{article}

\title{Fate of the Critical Line and Chiral Transition in
Finite-Temperature Lattice QCD with the Wilson Quark Action}

\author{S. Aoki\address{Institute of Physics, University of Tsukuba,
Tsukuba, Ibaraki 305, Japan}, A. Ukawa$^{\rm a}$, T. Umemura$^{\rm a}$}

\begin{document}

\begin{abstract}
Finite-temperature phase structure of lattice QCD with the Wilson quark
action is analyzed.
We show that the critical line at finite temperatures, defined
to be the line of vanishing pion screening mass, turns back toward
strong coupling, forming a cusp on the $(\beta, K)$ plane, and that
the line of thermal transition runs past the tip of the cusp without
touching the critical line.
Previous results are discussed in the light of our findings.
\end{abstract}

\maketitle
\section{Introduction}

Despite extensive effort over the years\cite{earlywork,qcdpaxtwo,milc}
the finite-temperature phase structure
of lattice QCD with dynamical Wilson quarks has not been fully clarified to
date.  The difficulty, originating from an explicit breaking of chiral
symmetry with the Wilson action,
arises from the question in what sense the concept of
the critical line of vanishing pion mass at zero temperature
extends to the finite-temperature system and how the line behaves in the space
of the inverse gauge coupling constant $\beta=6/g^2$ and the hopping parameter
$K$.   This is an issue of fundamental importance, not answered by previous
studies, which only examined the relation between the line of thermal
transition and the zero-temperature critical line.
In this article we report our resolution of the issue\cite{auu} based on
the idea of spontaneous breakdown of parity and flavor
symmetry\cite{aoki}.

\section{Phase diagram for two flavors}

\begin{figure}[t]
\centerline{\epsfxsize=74mm \epsfbox{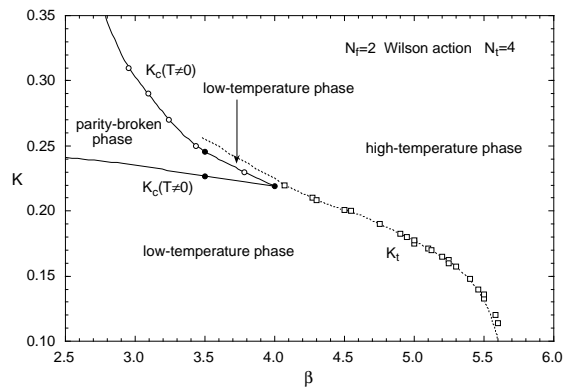}}
\vspace*{-10mm}
\caption{Phase diagram for $N_f=2$ with the Wilson quark action on an $N_t=4$
lattice.}
\vspace*{-5mm}
\label{fig:fig1}
\end{figure}

We define the critical line $K_c(\beta)$ for the finite-temperature system on a
lattice of a temporal lattice size $N_t$ by the vanishing of pion
screening mass
extracted from the pion propagator for a large spatial separation.
The line thus
defined represents a natural extension of the zero-temperature critical line
in that (i) it smoothly converges to the latter as
$N_t\to\infty$, and that (ii) it corresponds to a singularity of the
finite-temperature lattice QCD partition function.  Let us also define
$K_t(\beta)$ to be the line of thermal  transition, the thermal line.  With
these definitions the phase diagram we find for the case of $N_f=2$ flavors on
an $N_t=4$ lattice is presented in Fig.~\ref{fig:fig1}. The solid line is our
estimate of the critical line obtained by an interpolation of points where the
pion screening mass vanishes (solid circles) or the number of
conjugate gradient
iterations in hybrid Monte Carlo runs is estimated to diverge (open circles).
The dotted line is the thermal line with open squares representing its location
reported in previous simulations\cite{earlywork,qcdpaxtwo,milc}.

Let us emphasize the novel features of our phase diagram.
Contrary to the zero-temperature case, the critical line does not extend toward
arbitrarily weak coupling, but rather turns back toward strong coupling at
a finite
value of $\beta$, forming a cusp. The location of the critical line
before the turning point is generally shifted from that at zero temperature.
For
an $N_t=4$ lattice the shift is small, however.
We expect the tip of the cusp to be rounded if examined at a fine scale,
and that
there will be four more of such cusps in the entire $(\beta, K)$ plane.
The region
bounded by the critical line represents a phase in which parity and
flavor symmetry is
spontaneously broken\cite{aoki}.

We point out that the absence of the critical line
toward weak coupling provides a natural explanation of the fact, observed
in previous simulations\cite{qcdpaxtwo,milc},
that physical quantities in the high-temperature
phase vary smoothly across the zero-temperature critical line.

Another important feature of our phase diagram is that the thermal line
runs past the tip of the cusp of the critical line and continues toward larger
values  of $K$.  In other words the low-temperature phase goes around the
cusp and extends into the region bordered by the upper part of the critical
line.
This contrasts with the conclusion of the QCDPAX Collaboration that the
two lines meet at $\beta=3.9-4.0$\cite{qcdpaxtwo}.
This point is discussed later.

\section{Evidence for the phase diagram}
\vspace*{-2mm}
\subsection{Gross-Neveu model in two dimensions}

\begin{figure}
\centerline{\epsfxsize=70mm \epsfbox{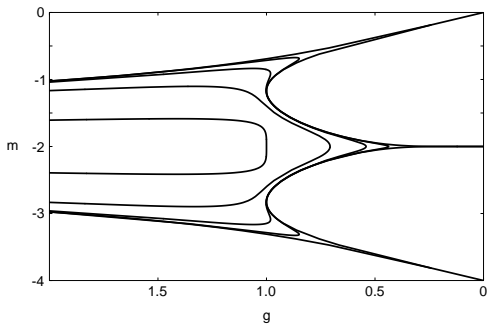}}
\vspace*{-10mm}
\caption{Critical line for the Gross-Neveu model.  Lines correspond to temporal
lattice size $N_t=2,4,8,16$ and $\infty$ from inside to outside.}
\label{fig:fig2} \vspace*{-5mm}
\end{figure}

An instructive hint on the nature of the critical line is provided by the
two-dimensional Gross-Neveu model formulated with the Wilson quark
action\cite{eguchinakayama}. In the large $N$ limit the critical line of this
model can be calculated from the saddle point equations for the pion mass.  The
result is shown in  Fig.~\ref{fig:fig2} on the plane of the bare coupling
constant $g$ and bare quark mass $m=1/2K-2$, where the
temporal lattice size is taken to be $N_t=2, 4, 8, 16$ and $\infty$ from inside
to outside.

At zero temperature there are
multiple critical lines forming three cusps at $g=0$.  Outside of the
critical line is the normal phase, while inside is
the parity-broken phase characterized by
$\langle\overline{\psi}\gamma_5\psi\rangle\ne 0$. Pion is
interpreted as the zero mode of the second-order transition separating the two
phases\cite{aoki}.  The conventional continuum limit is taken near
$(g,m)=(0,0)$ along the
upper critical line.  The existence of two additional cusps arises from an
exchange of doublers and physical fermion as $m$ is varied.

At finite temperatures the cusps move away from the
weak-coupling limit $g=0$.  As a result the critical line forms
a single continuous
line.

\begin{figure}
\centerline{\epsfxsize=74mm \epsfbox{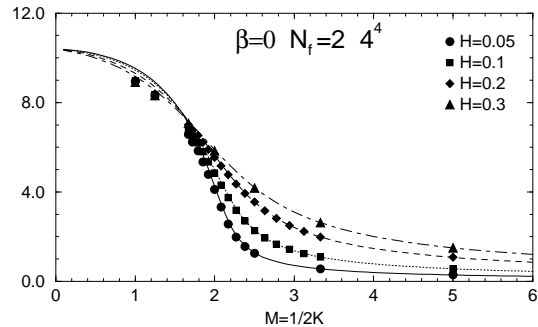}}
\vspace*{-10mm}
\caption{Hybrid Monte Carlo results for parity-flavor order parameter
$\langle\overline{\psi}i\gamma_5\tau_3\psi\rangle$ for
the $N_f=2$ case at $\beta=0$ on a $4^4$ lattice as compared to
predictions of  $1/N_c$ expansion (drawn by lines).}
\label{fig:fig3}
\vspace*{-5mm}
\end{figure}

\subsection{Zero-temperature simulations}

The chiral properties of the Gross-Neveu model are similar to those
expected for
lattice QCD.  This leads us to anticipate that the structure and
characterization
of the critical line are also similar except
that there will be five cusps for QCD due to an increase of dimensions from
two to four.

The existence of a parity broken phase in QCD was originally suggested by an
analytical calculation in the strong coupling limit in the
$1/N_c$ expansion\cite{aoki}.
In Fig.~\ref{fig:fig3} we show a comparison of the analytical prediction and
our hybrid Monte Carlo results for the parity-flavor order parameter
$\langle\overline{\psi}i\gamma_5\tau_3\psi\rangle$ for the $N_f=2$ case
in the presence
of a symmetry-breaking external field $H$.
A good agreement between the two results
supports the conclusion of the large $N_c$ analysis that parity
and flavor symmetry become spontaneously broken for $K\geq 1/4$ at $\beta=0$.

\begin{figure}
\centerline{\epsfxsize=74mm \epsfbox{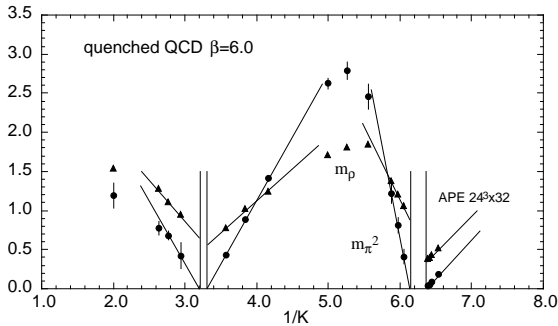}}
\vspace*{-10mm}
\caption{Zero-temperature $\pi$ and $\rho$ mass at $\beta=6.0$
in quenched QCD.}
\label{fig:fig4}
\vspace*{-5mm}
\end{figure}

In Fig.~\ref{fig:fig4} we plot the zero-temperature $\pi$ and $\rho$ masses at
$\beta=6.0$ in quenched QCD as a function of $1/K$.  The right-most set of data
points are taken from the APE Collaboration\cite{ape}, while the rest of data
are obtained with $10-20$ configurations on a $10^3\times 20$ lattice.  In
addition to the conventional critical line at $1/K=6.3662$, we observe clear
indictions for three more values of $K_c$.  While the fourth value expected
toward $1/K\to 0$ is yet to be confirmed, this provides evidence that there are
multiple critical lines toward  weak coupling in QCD.

\subsection{Finite-temperature simulations}

In order to determine the shape of the finite-temperature critical line,
we have carried out hybrid Monte Carlo simulations for the $N_f=2$ system on an
$8^3\times 4$ lattice in the region $3.0 \leq\beta\leq 5.3$
and $0.15\leq K\leq 0.31$.

At $\beta=3.5$ results for the pion screening mass demonstrate
that there exists
a second critical value at $K_c=0.2454$ beyond the
conventional value at $K_c=0.2267$ (see Fig.~2(a) in Ref.~\cite{auu}).
The behavior of the pion mass is
similar to that around $1/K\approx 6.3$ in Fig.~\ref{fig:fig4}.
A parallel analysis at $\beta=4.0$ shows that the gap between the two
critical values is either extremely narrow or disappears just below
$\beta=4.0$
(Fig.~2(b) in Ref.~\cite{auu}).
Since previous simulations\cite{earlywork,qcdpaxtwo,milc} found no
singular behavior beyond the thermal line down to $\beta=4.3$\cite{qcdpaxtwo},
we conclude that the critical line turns back toward strong coupling in the
vicinity of $\beta\approx 4.0$.

We expect the number of conjugate gradient iterations $n_{\rm CG}$
for evaluating
quark propagators in hybrid Monte Carlo runs to diverge toward
the critical line.
Extrapolating $1/n_{\rm CG}$ linearly in $\beta$ at a fixed $K$,
we estimate the
location of the critical line for several values of $K$. The results,
together with
those from the pion mass, are smoothly connected, forming a cusp as is shown in
Fig.~\ref{fig:fig1}.

A characteristic feature of the high temperature phase is that the
pion screening mass takes a finite value.  Such a phase can not include the
critical line where the screening mass vanishes.
Thus the region close to the entire critical line has to be in the
low temperature phase.  The validity of this argument is confirmed through
measurements of the quark and gluon entropy and the Polyakov line.  Results at
$\beta=3.5$ show that these quantities decrease toward the second
critical value
when it is approached from larger values of $K$, and become comparable in
magnitude to those below the conventional critical value which is in the low
temperature phase (see Fig.~3(a) in Ref.~\cite{auu}).

This means that the thermal line can not cross the critical line.
Since the two lines are known to come close around
$\beta\approx 4.0$\cite{qcdpaxtwo}, we conclude that the thermal line
runs past the tip of the cusp and continues toward larger values of $K$.
Our results for
thermodynamic quantities taken around the tip of the
cusp at $\beta\approx 4.0$ (Fig.~3(b) in Ref.~\cite{auu}) and those in the
literature\cite{qcdpaxtwo,milc} are consistent with this conclusion.

\vspace*{-3mm}
\section{Discussions}
\vspace*{-2mm}
\subsection{Comparison with previous results}

A naive expectation for the chiral transition with dynamical Wilson
quarks is that the thermal line would intersect the zero-temperature
critical line, which would naturally define the point of chiral transition
in that it would be the point of vanishing pion mass separating the low- and
high-temperature phases. Early work\cite{earlywork} guided by this idea
failed to find the crossing point.  More recently the QCDPAX
Collaboration concluded that the crossing point with the expected properties
is located at $\beta\approx 3.9-4.0$ for the $N_f=2$ system
on an $N_t=4$ lattice\cite{qcdpaxtwo}.

However, the actual structure differs from their
conclusion in one crucial aspect: while the thermal line does cross
the zero-temperature critical line since the latter continues toward
the weak coupling limit,
the pion screening mass does not vanish at this crossing point.
In fact a unique point marking chiral transition as naively expected does
not exist in our phase diagram.  The whole region around the tip of the
cusp and the thermal line is relevant for understanding the chiral
properties of the thermal transition.

\subsection{Dependence on quark action}

The key ingredients of our analyses are the properties of the Wilson quark
action
that massless free quarks appear at multiple values of the hopping parameter
and
that the action has parity-flavor symmetry.  We therefore expect the
qualitative
structure of our phase digram to remain valid for a more general form of
Wilson-type actions including the clover action.

We remark that our phase diagram does not explain the metastability
signals at $\beta\approx 5.0$ found by the MILC Collaboration\cite{milc}.  This
phenomenon is probably a lattice artifact closely related to the specific
form of
the Wilson plaquette or quark action\cite{qcdpaximp}.

\subsection{Dependence on $N_f$}

We expect our phase diagram
to apply also for a larger number of flavors, at least up to $N_f=6$,
except for
one difference. While simulation results strongly indicate that the thermal
transition for the $N_f=2$ case is a continuous crossover and not a true phase
transition,  first-order signals are found for for
the case of $N_f=3$ and 6 close to the zero-temperature critical
line\cite{qcdpaxtwo}.   This implies that the thermal line turns into a line of
a first-order phase transition while running close to the tip of  the cusp of
the finite-temperature critical line for these cases.

For $N_f\geq 7$ the QCDPAX Collaboration reported that the critical line
disappears even in the limit of strong coupling.  Understanding this phenomenon
by an extension of our line of analysis is an interesting problem left for
future studies.

\subsection{Continuum limit}

When we increase the temporal lattice size $N_t$, we expect both the
thermal line and the cusp of the critical line to move toward larger $\beta$.
The distance between the thermal line and the tip of the cusp
will diminish, probably as $O(a)\approx O(1/N_t)$.
To elucidate the properties of chiral transition in the continuum limit,
one should measure thermodynamic
observables across the thermal line in the region around the tip of the cusp
and examine how they vary as a function of $N_t$.

An important point to note is that the $N_f=2$ transition is a continuous
crossover for finite values of $N_t$.
A second-order
chiral phase transition, as suggested by continuum sigma model analyses for
this case, would emerge only in the continuum limit $N_t\to\infty$.
\vspace*{2mm}

We thank Y. Iwasaki, K. Kanaya and T. Yoshi\'e for
useful discussions.
This work is supported in part by the Grants-in-Aid of
the Ministry of Education (Nos. 04NP0701, 06640372).


\begin{thebibliography}{9}

\bibitem{earlywork}M. Fukugita, S. Ohta and A. Ukawa,
Phys. Rev. Lett. 57 (1986) 1974; A. Ukawa, Nucl. Phys. B(Proc.
Suppl.)9 (1990) 463; R. Gupta {\it et
al.,} Phys. Rev. D40 (1989) 2072; K. M. Bitar {\it et al.,}
Phys. Rev. D43 (1991) 2396.

\bibitem{qcdpaxtwo}Y. Iwasaki {\it et al.}, Phys. Rev. Lett. 67 (1991) 1491,
69 (1992) 21; Nucl. Phys. B(Proc. Suppl.)30 (1993) 327, 34 (1994) 314;
hep-lat/9504019, 9505017.

\bibitem{milc}C. Bernard {\it et al.}, Phys. Rev. D46 (1992) 4741,
49 (1994) 3574, 50 (1994) 3377.

\bibitem{auu}S. Aoki, A. Ukawa and T. Umemura, hep-lat/9508008.

\bibitem{aoki}S. Aoki, Phys. Rev. D30 (1984) 2653;
Phys. Rev. Lett. 57 (1986) 3136; Nucl. Phys. B314
(1989) 79;
hep-lat/9509008.

\bibitem{eguchinakayama}T. Eguchi and R. Nakayama, Phys. Lett. 126B (1983) 89.

\bibitem{ape}S. Cabasino {\it et al.,} Phys. Lett. B258 (1991) 195.

\bibitem{qcdpaximp}Y. Iwasaki {\it et al.,} Nucl. Phys. B(Proc. Suppl.) 42
(1995) 502; in these proceedings.

\end{thebibliography}
\end{document}